\def\year{2024}\relax
\documentclass[letterpaper]{article} % DO NOT CHANGE THIS
\usepackage{aaai22}  % DO NOT CHANGE THIS
\usepackage{times}  % DO NOT CHANGE THIS
\usepackage{helvet}  % DO NOT CHANGE THIS
\usepackage{courier}  % DO NOT CHANGE THIS
\usepackage[hyphens]{url}  % DO NOT CHANGE THIS
\usepackage{graphicx} % DO NOT CHANGE THIS
\usepackage{natbib}  % DO NOT CHANGE THIS AND DO NOT ADD ANY OPTIONS TO IT
\usepackage{caption} % DO NOT CHANGE THIS AND DO NOT ADD ANY OPTIONS TO IT
\DeclareCaptionStyle{ruled}{labelfont=normalfont,labelsep=colon,strut=off} % DO NOT CHANGE THIS
\usepackage{algorithm}
\usepackage{algorithmic}
\newtheorem{theorem}{Definition}

\usepackage{newfloat}
\usepackage{listings}
\floatstyle{ruled}
\newfloat{listing}{tb}{lst}{}
\floatname{listing}{Listing}

\usepackage{natbib}  % DO NOT CHANGE THIS AND DO NOT ADD ANY OPTIONS TO IT

\usepackage{longtable}
\usepackage{graphicx}
\usepackage{booktabs}

\usepackage{subcaption}
\usepackage{todonotes}

\usepackage{xcolor}

\newcommand{\vknote}[1]{\textcolor{teal}{VK: #1}}
\renewcommand{\vknote}[1]{}
\newcommand{\xhdr}[1]{\paragraph{\bf {#1}}}

\begin{document}
\title{What's in a Niche? Migration Patterns in Online Communities}

\author{Katherine Van Koevering,\textsuperscript{1}
Meryl Ye,\textsuperscript{1}
Jon Kleinberg,\textsuperscript{1}\\
\textsuperscript{1}{Cornell University}\\
kav64@cornell.edu,
may43@cornell.edu, 
kleinberg@cornell.edu}

\maketitle

\begin{abstract}
Broad topics in online platforms represent a type of meso-scale between individual user-defined communities and the whole platform; they typically consist of  related communities that address different facets of a shared topic. 
Users often engage with the topic by moving among the communities within a single category. 
We find that there are strong regularities in the aggregate pattern of user migration, in that the communities comprising a topic can be ordered in a partial order such that there is more migration in the direction defined by the partial order than against it.
Ordered along this overall direction, we find that communities in aggregate become smaller, less toxic, and more linguistically distinctive, suggesting a picture consistent with specialization.
We study directions defined not just in the movement of users but also by the movement of URLs and by the direction of mentions from one community to another; each of these produces a consistent direction, but the directions all differ from each other.

We show how, collectively, these distinct trends help organize the structure of large online topics and compare our findings across both Reddit and Wikipedia and in simulations.

\end{abstract}

\section{Introduction}

Social media is a remarkably dynamic landscape. Users, information, and even platforms are constantly on the move. Users' interests and interactions evolve and as they do the users move from one corner of the internet to another. Studying these user migrations has been popular \cite{kumar2011understanding, newell2016user, davies2021multi} especially in recent years, where issues around conspiracies and `rabbitholes' has sparked renewed interest in understanding where and how users move into increasingly extreme communities \cite{fernandez2018understanding, wang2023identifying}. 
 
As a factor of these migrations, many online sites follow an organizational scheme composed of numerous groups or communities, each structured around a topic of shared interest. Many of the earliest forums and message boards on the Internet had this design, and it is a central feature for several of the most active sites today, such as Reddit. This type of structure based on shared-interest communities is also a feature of how people pursue interests in the offline world, with clubs and organizations bringing people together \cite{glorieux2010search, gelber1999hobbies}. 

Moreover, the internet is a big place. On large sites, it is rare for there to be one singular community dedicated to any broad interest. Instead, we find a collection of communities each with a different theme, but all related to this interest \cite{tan2015all, teblunthuis2022no, teblunthuis2022identifying}. For instance, instead of one community discussing US politics, one might find several communities at different points of the political spectrum, or reflecting different levels of involvement.

We pursue the question of user migration in this context. In particular, does a single user become active in multiple communities on the same topical category? And if so, is there a directionality to their movement---do the communities on the same topic form a kind of partial ordering or gradient exhibiting more motion in the direction of the gradient than against it? We cannot expect such properties to fully express the idiosyncratic decisions that every user makes on a large site, but we can look for evidence of aggregate effects, much as we might for directionality in the physical world: a collection of %seed pods floating in the air, or 
twigs in a body of water, might either be moving uniformly in all directions, or they might have a general drift one way, indicating a current. %breeze or a current in the water. 

We study these questions across a range of social media topics, each of which contains multiple communities.  Within each of the topics we study, we find strong evidence to support the existence of these directional migrations. 
We focus our analysis on topics on Reddit, where the subreddits on a particular topic form the related communities, and we consider analogues on Wikipedia as well.

\begin{figure}
    \centering
    \includegraphics[width=0.45\textwidth]{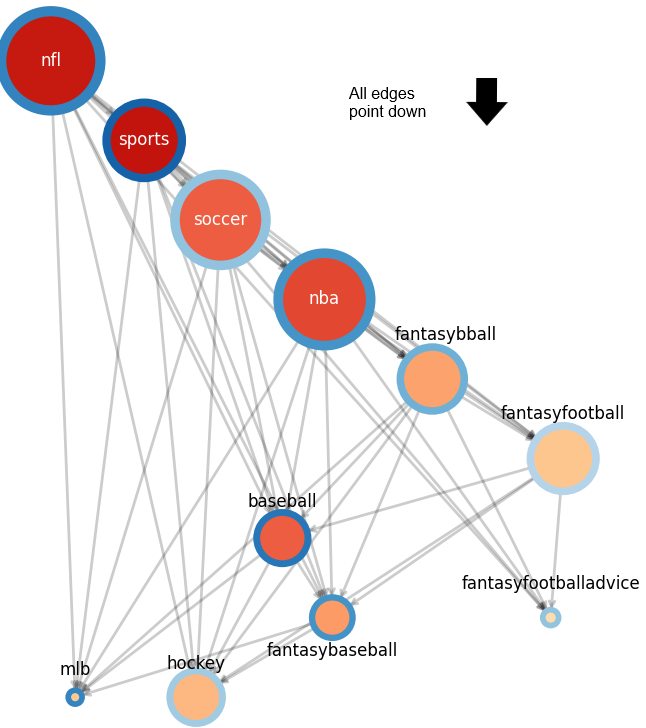}
    \caption{The user gradient for sports communities on Reddit, one of the topics we study, demonstrates how three of our measures (size, toxicity, and linguistic distinctiveness) are related to the aggregate progression of users along a gradient - communities become smaller, less toxic, and more distinct. Edges proceed downward. The size of nodes is a logarithmic scale of the size of communities. The inner color of the nodes represents the toxicity of the community, where darker nodes are more toxic. The outer color of the node represents linguistic distinctiveness, where darker edges represent less distinctive communities.}
    \label{fig:tox}
\end{figure}

The consistent directionality of user migration then raises the question of what factors are influencing the direction. 
There are two common narratives of the internet that are often used informally to describe user migration: radicalization, in which users move to communities with increasingly extreme behavior \cite{fernandez2018understanding}, and specialization, in which users move to communities with increasingly focused, niche interests  \cite{waller2021quantifying}.
In our analysis, 
we find that, overall, users tend to move to smaller communities that linguistic measures evaluate as lower in toxicity but higher in lexical distinctiveness (Fig~\ref{fig:tox}).
We find that this change in the content of communities over users' aggregate direction of movement provides an interesting perspective on these two narratives in our context: the aggregate decrease in toxicity as a user moves between communities works against the narrative of radicalization, whereas the increase in linguistic distinctiveness is consistent with specialization.
We emphasize, of course, that each individual user will have their own motivations in moving between communities, but the contrasts stands out at the level of the overall user population.

In addition to user migration, we also investigate information migration \cite{jin2023predicting} in a similar fashion using two measures: posting URLs and comments with direct references to other communities. Specifically, we find three distinct types of trends that emerge from this analysis.
The first two types of trends are (i) an {\em individual (user) migration} of the type described above, defined by consistent patterns in how people move between communities within a topical category; and (ii) an {\em informational migration}, defined by the analogous movement of discrete pieces of information between communities, encoded by following URLs as they are subsequently posted in various communities. Both of these migrations are marked by temporal patterns of engagement.

Our third type is (iii) a {\em referential directionality}, which looks at references, or {\em mentions}, from one community to another, asking which communities talk about others more.
In this case, we ask whether it is possible to define an overall direction based on the asymmetry in how much different communities mention each other.
We find that all three measures produce strong trends, but are distinct in their orderings and influences. Finally, we compare our findings to similar analysis of Wikipedia editors and the migrations extant in that space, as well as simulated networks.

In the course of this analysis, we discover a number of interesting factors that augment our understanding of online migration patterns. For instance, users have a tendency to move wholly from one community to another, rather than straddle many communities simultaneously within a category. However, when users talk about other communities within a category, the sentiment trends negative.

\subsection{Additional Related Work}
Many prior works have considered categories of topically related communities.
One strand of research has studied {\em spinoff communities} \cite{hessel2016science, zhu2014selecting}, focusing on the creation (and eventual death) of highly related shared-interest communities. Additional work has turned the focus to what makes these communities different and how they evolve alongside each other \cite{rajadesingan2020quick, zhang2021understanding, mcthenia2021organizing}.  This includes instances of how communities interact and respond to each other outside a user perspective. \cite{teblunthuis2022identifying, teblunthuis2022no}.

Understanding the movement of users and information through social networks has been a richly studied topic, although much analysis in this literature is not directly relevant to us here. Instead, we specifically note several papers that focus on user life cycles \cite{yang2010activity, tan2015all, rowe2013mining}, and more recent work on user embeddings \cite{waller2021quantifying}. All of these works attempt to understand or explain why users participate in certain online communities, and how these groups evolve over time (and how users might choose to join, participate, or leave). Some of this work, including Tan et al. \cite{tan2015all} Zhang et al \cite{zhang2021understanding}, and Rowe \cite{rowe2013mining} focus on instances of users participating in shared interest groups and highly related communities, principally from a user perspective. Finally, a recent paper established URL information flow using YouTube links on Reddit as an interesting and complex problem \cite{jin2023predicting}.

\nocite{myers1976group}

\section{Data Description}\label{sec:Data}
The data used in this research is collected from Reddit, a popular social media platform, using Pushshift, a nearly-comprehensive archive of Reddit (which until recently was publicly available). This discussion-based platform has a large number of user-created individual forums known as subreddits. Each subreddit has an explicitly stated topic, usually related to its name that is chosen by the founder. This clear definition of communities and topics makes Reddit ideal for studying community interaction. Due to this, Reddit is also an extraordinarily well-studied platform \cite{medvedev2019anatomy, proferes2021studying, jamnik2017use}, with extensive research into Reddit behavior \cite{amaya2021new}, user demographics \cite{tigunova2020reddust}, user motivations \cite{bogers2014social, moore2017redditors} communities \cite{monti2023evidence, soliman2019characterization}, information sharing \cite{de2021no, zimmer2018evaluation}, and more \cite{baumgartner2020pushshift}. As such, we refer to these papers, and many others on similar subjects, for detailed analysis of Reddit, Reddit demographics, and description of Reddit as a data source and social media platform. 

Reddit also holds a unique role on the internet. Unlike other large social media platforms such as Facebook, Twitter, Instagram, TikTok or YouTube, Reddit represents a socially mediated form of expertise. Reddit users do not generally engage in "following" particular users the way one might on most other social media, nor is there a "friend" mechanic - this centers activity on Reddit around the information presented rather than the users themselves. It is primarily a source for finding information and it is common for users to interact with other invested, niche, and expert users and engage with posts or comments months or even years old. It is also community focused in a way that other platforms are not, with its structuring around themed communities giving distinct boundaries for topics, cultures, and users. Together, this presents Reddit as a prime location to identify and study these themed community groups, where the communities are highly distinct and the themes influential.

The selection of our communities is an important step in our analysis that requires detailed explanation\footnote{Code and data will be made available upon publication}. To begin, we selected thirteen categories of topics that represent a wide array of users and have many different yet related communities on Reddit. Among these is politics, which we chose for comparative purposes. The rest were selected according to three principles: (i) representation of multiple levels of specificity (e.g., the category 'sports' is considerably more broad than the category 'Indoor Plants'), (ii) representation of various demographics (i.e., choose categories that have a good chance of representing diverse demographics, particularly in terms of age, race, and gender),  (iii) representation of various degrees of politicization (political radicalization is a well-studied phenomenon that should not strongly influence all of our categories). See the appendix for more detail on community choice.

\section{User Migration}

\begin{table*}
    \centering
    \begin{tabular}{lcrr}
    \toprule
    category & \# communities & \# posts/comments & \# users\\
    \midrule
         Computer Science (CS)& 19 &54,304,503 &1,800,653\\
         Camping& 13 &13,690,588 &620,178\\
         Crafting& 16 &24,658,755 &1,383,781\\
         Early Retirement& 14 &56,866,540 &5,938,592\\
         Fitness& 20 &5,4256,622 &3,483,201\\
         Indoor Plants& 16 &13,302,290 &1,221,205\\
         Makeup& 19 &58,141,329 &1,701,287\\
         PC Building& 14 &82,819,446 &3,426,274\\
         Politics& 13 &308,943,275 &91,915,842\\
         Sports&16 &86,770,040 &8,610,698\\
         Survival&14 &17,298,659 &605,557\\
         Veganism&11 &15,591,562 &977,653\\
    \bottomrule
    \end{tabular}
    \vspace{2mm}
    \caption{Categories, number of associated communities, and the total number of posts/comments and users collected for each.}
    \label{tab:cats}
\end{table*}

Within a category, users must choose which communities to participate in and when, but what choices they make and what factors influence those choices is unclear. Understanding user flow across the Internet has been a long-standing goal of Web research over many years \cite{hessel2016science,huberman1998strong,rowe2013mining,tan2015all,yang2010activity}, and while a number of influences have been proposed, we are far from truly understanding the elements that influence our online lives.

User behavior on Reddit is a well-studied phenomenon and we refer to the myriad of prior work for more detailed descriptions \cite{amaya2021new, tigunova2020reddust, bogers2014social, choi2015characterizing, newell2016user, thukral2018analyzing, glenski2017consumers}. It is sufficient to say that users follow previously established behavioral patterns---exponentially distributed amounts of activity, bursty activity, and exponentially distributed number of communities.

\subsection{Defining User Migration}
\begin{figure}
    \centering
    \includegraphics[width=0.45\textwidth]{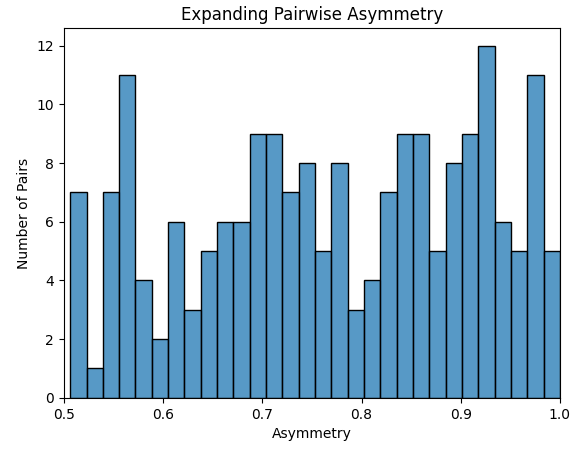}
    \caption{Histogram of the asymmetry scores for all pairs of communities.  Adjusting $k$ has little effect. Note that the average asymmetry of user directions is far more  than would be expected with random movement (with a p-value of nearly 0), and random movement would produce an exponential distribution rather than a more uniform distribution.}
    \label{fig:users_ass_hist} 
\end{figure}

Given that users are migrating between communities, this raises the question of whether all users are moving in the same way or if they move differently. As a first step, we need a way to formalize a user's association with a given community; in particular, it is useful for the analysis to require a threshold of involvement that may be more than a single post, but which is permissive enough to include a wide range of activities in the community.
In particular, we say that a user has {\em entered} a community when they first exceed our definitional threshold for involvement. We  have explored a wide range of definitions for this process of a user entering a community, and they give markedly similar results.
As such, we compare the following two possible definitions:

\begin{theorem}[Thresholding]
A user must post/comment in a community at least $k$ times, and enters on the $k$th post/comment
\end{theorem}
\begin{theorem}[Expanding]
A user must post/comment in a new community at least $k$ times without posting/commenting in any other new communities, and enters on first post/comment in this sequence.

\end{theorem}

The most obvious definition for entering a community may be the thresholding definition. However, this definition produces arbitrary behavior when, for instance, a user is alternating between two new communities. Thus using this definition produces a noisier picture. The expanding definition, on the other hand, is much more stringent. It reduces the noise and focuses our analysis on users who make more definitive moves to a new community. We choose the expanding definition for the rest of our analysis, but the thresholding definition yields nearly identical, albeit noisier, results. 

We also impose a minimum number of posts/comments in order for a user to be viewed as participating in a communities, rather than a transient contribution; for our purposes, we require at least $k=20$ posts/comments, although raising and lowering $k$ has minimal effect on results. One more novel aspect of user behavior is how users change their behavioral patterns after entering a new community - users tend to either remain primarily tied to their original community or switch completely, with only a small percentage of users participating in both relatively equally. This emphasizes the temporal aspect of user participation in a community and lends credence to our expanding definition.

\subsection{Gradients of User Migration}
We now consider pairs of communities:
for a given pair of communities $(A,B)$, we say that the {\em orientation} of $(A,B)$ is directed from $A$ to $B$ if a majority of users associated with both enter $A$ before $B$; the orientation is directed from $B$ to $A$ if a majority enter $B$ before $A$;
and the orientation is {\em void} if the number of pairs who enter both is below some specified threshold. 
We define the {\em asymmetry} of the orientation of $(A,B)$ to be the fraction of users that move in the majority direction. For instance, we say that the orientation of $(nba, fantasybball)$ is from $nba$ to $fantasybball$ since most users who are part of both communities enter $nba$ first.
\begin{theorem}[Orientation \& Asymmetry]
    Let $A$ and $B$ be two communities s.t. some $n \geq k$ number of users have, by our definition, entered both communities. Let the number of users who entered $A$ before $B$ be $a$ and the number of users who entered $B$ before $A$ be $b$ s.t. $a + b = n$. Let $a > b$. 
    
    Then, we say the \textbf{orientation} is $(A,B)$ and the \textbf{asymmetry} is $\frac{a}{n}$.
\end{theorem}

For the communities within a category, we can form a directed graph whose nodes are the communities, and where there is a directed edge from $A$ to $B$ if the orientation of $(A,B)$ is from $A$ to $B$. We say that a set of orientations on pairs of communities forms a {\em gradient} if this directed graph is acyclic. 

\begin{theorem}[Gradient]
    For each pair of communities $A$, $B$, if there is an orientation $(A,B)$ then let there be two nodes $A$ and $B$ with a directed edge between them from $A$ to $B$. This graph, when acyclic, represents a partial ordering on all communities. We refer to this partial ordering as the \textbf{gradient}.
\end{theorem}

Given random movement, as describe in the next section, we would not expect any patterns in this network. However, surprisingly, across all 13 of the categories we study, the graphs of strong edges are acyclic. This means for all our categories, the set of orientations on the pairs of communities forms a gradient, indicating a consistent direction of user movement through the category. This is extraordinarily unlikely to happen at random, as described below, and indicates there are likely strong influences on the direction of movement between communities. We will spend much of the rest of this paper exploring these influences.

\begin{figure}
    \centering
    \includegraphics[width=0.45\textwidth]{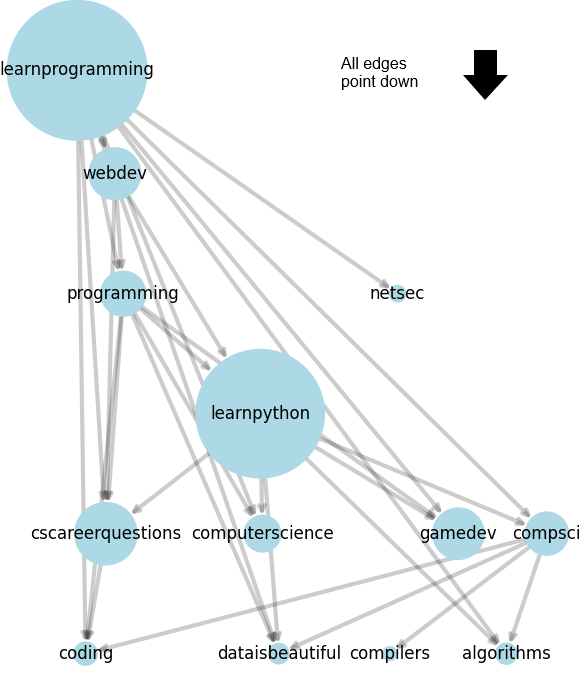}
    \includegraphics[width=0.45\textwidth]{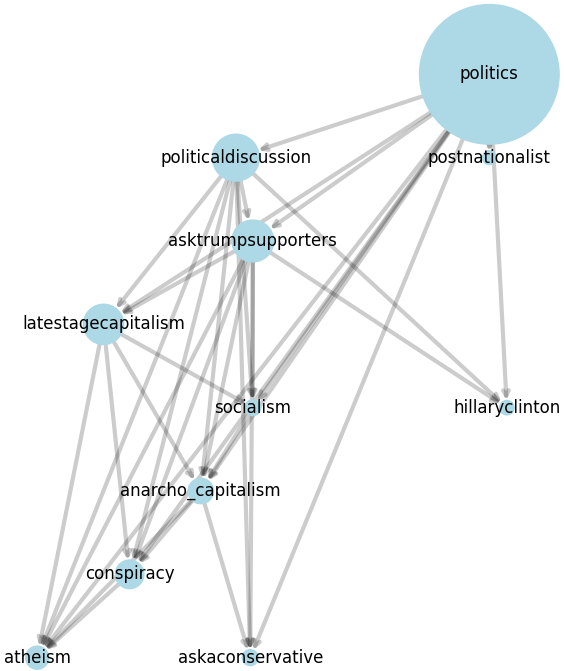}

    \caption{Small graphs representing the user gradient for politics and CS, where all edges move downwards. Note that not all communities in each category appear in these graphs. Those that do not, do not have strong enough user overlap with any other community ($<20$ users by our definition). For CS, we remove all edges of less than 0.75 asymmetry to remain acyclic, our only thresholding for the use gradient.}
    \label{fig:thresh}
\end{figure}

In these gradients, a large enough fraction of community pairs have edges, such that
there are relatively few distinct topological orderings of the graphs: they are often essentially linear (Fig~\ref{fig:thresh}). Additionally, the fact that every community present in the graph is very connected within it, despite not all communities being represented, suggests we have found a natural boundary of the category.

\subsection{Null Models}
We compare our small acyclic graphs with two null models - randomizing edge directions and randomizing edge placement. In the first, we randomly re-assign the edge direction in a graph. We do this many times, and ask how many result in an acylcic graph. For the second, we create a new graph with the same number of edges and nodes, where the edges are placed uniformly at random. Again, we repeat this process and count the number of acyclic graphs. For most of our categories, we find significantly few acylic graphs (less than 1\%) Tab~\ref{tab:null}. The exceptions are the few categories with very few edges. This suggests that it is extremely unlikely for the acyclic graphs we see here to have happened at random. As a comparison, less than $4.1*10^{-4}$\% of all possible directed graphs of 10 nodes are acyclic.

\begin{table}
    \centering
    \resizebox{\linewidth}{!}{\begin{tabular}{lccc}
    \toprule
      Category&  \# communities & \#Edges & \ Fraction Randomly Acyclic\\
         \midrule
         Computer Science (CS)& 13 &28 & $<0.01$**\\
         Camping& 7 &11 &0.21\\
         Crafting& 10 &28 & $<.01$**\\
         Early Retirement& 10 &12 &0.39\\
         Fitness& 10 &23 & 0.01*\\
         Indoor Plants& 9 &24 & $<0.01$**\\
         Makeup& 6 & 10 & 0.19\\
         PC Building& 11 &43 & $<0.01$**\\
         Politics& 11 &34 & $<0.01$**\\
         Sports&11 &48 & $<0.01$**\\
         Survival&4 &6 &0.34\\
         Veganism&6 &8 &0.34\\
    \bottomrule
    \end{tabular}}
    \caption{The size of the acyclic graphs, and the fraction of random graphs that have acyclicity under the edge-swap model.}
    \label{tab:null}
\end{table}

\subsection{Extremists and Specialists}

\begin{table}[]
    \centering
\begin{tabular}{lll}
\toprule
  \# users &  Decreases ** \\
  total activity & Decreases* \\
  activity per user & Increases \\
  toxicity &  Decreases** \\
  distinctiveness &  Increases**\\
\bottomrule
\end{tabular}
\vspace{1mm}
\caption{Binomial tests comparing source and target communities for all edges in user movement graphs. * is p-value $<.05$, ** is p-value $<.01$. We measure toxicity with the Google Persepctive API.}
\label{tab:expanding_pairs}
\end{table}

Given the existence of these gradients, it is natural to ask what is informing their ordering. The mechanisms behind this pattern of user migration (and, indeed, informational and referential directionality as well) is likely complex, but we address some aspects of the mechanism here. One simple test for relevance is to ask for correlation of various measures with the edges of our graphs - asking if there is a significant difference in measures between communities $A$ and $B$ where we have a trend of movement from $A$ to $B$. By this analysis, we find three measures with any significant relevance: size, toxicity, and distinctiveness. These three measures are statistically independent and relevant to two competing narratives of the Internet: radicalisation and specialization.

\xhdr{Size} Size can be measured in a variety of ways - number of distinct users, number of distinct posts/comments, number of distinct URL's, etc. Unsurprisingly, these measures are all highly correlated with each other. If we use number of users as our proxy, we see strong correlation with our gradient. That is, generally, users tend to move from large to small communities. 

\xhdr{Toxicity} Toxicity is another measure. While one might expect users to be drawn into a `rabbithole' of increasingly toxicity, we actually find the opposite. There is a significant correlation with decreasing toxicity and user gradients. Users tend to move from more toxic to less toxic communities.

\xhdr{Distinctiveness} Finally, we investigate distinctiveness. That is, we ask how different the language used in a subreddit is from the language used in other subreddits. We test this using tfidf and a custom term frequency measure, but both produce similar results. Users generally move to communities with more distinct language. In practice, this distinctiveness tends to represent increased usage of technical terminology within the category.

These three measures together bring to mind three competing narratives of  the internet. The first narrative, radicalization, harkens to research in political radicalization and suggests that over time, internet denizens tend to move to more radical and extreme spaces \cite{wang2023identifying}. A second narrative, specialization, references human behavior offline in that users become more specialized in niche or advanced topics within a category \cite{waller2019generalists, mcauley2013amateurs}. 

Both behaviors exist and both could plausibly be influential in these gradient orderings. If we were to posit that these user gradients mirror the radicalization seen in political spheres, the extreme communities narrative, one might expect increased engagement or increased toxicity. Instead, we find that the amount of engagement per user does not change and toxicity decreases. If we were to instead suggest that that the narrative of specialization informs the ordering, we might expect size to decrease, but distinctiveness to increase, with agnosticism on toxicity and engagement - which we do see.

\section{Informational and Referential Movement}
The definitions for movements and gradients defined above do not only apply to users. We can also apply them to information and references. In this case, we use URLs as an example of information movement. If a URL is first posted in one community and then later in another, we can view this similarly to a user entering these communities. This gives us a rough proxy for information movement. On the other hand, references are quite different. We say a community \textbf{mentions} another community if the text "r/$community$" exists in any post or comment. Unlike users and URLs, there is no real concept of movement, but we retain the directed pairs structure.

While user migration has analogues in prior research, albeit with a variety of definitions, URL movement and mentions are more novel under any formalism. As such, it is useful to develop a better understanding of the underlying nature of these measures. We begin by evaluating the size of each measure. The sizes are closely correlated---larger subreddits also use more URLs and more mentions. Note that with the differing scales, however, mentions are by far the most common of the measures.

\subsection{URL \& Mention Behavior}
URLs and mentions have an additional structure in that they are references outside of the community. URLs are a representation of retrieving additional information and bringing that to an ongoing conversation. A flow of URLs is akin to a flow of information. Thus, it is inherently interesting to understand where this information is coming from (Tab~\ref{tab:perc_reddit}). In fact, a large percentage of these URLs reference Reddit (and we include these as `mentions' in our later analysis). Additionally, not only do URLs link to other Reddit communities, they may also link to another post or comment in the same community. We see wide variation in these. 

Since URLs are pointing at specific conversations on Reddit, this implies that some categories may have a stronger `memory' in that they review previous material more often. It also implies that some categories share information among themselves more than other categories do.

URLs, much like users, can be posted in a community multiple times. The vast majority of URLs, however, are only posted once. In fact, a large percentage of URLs that appear in many communities (particularly cross-categories) are related to bot accounts or Reddit settings (e.x., links to spam reporting, summarizer bots, auto-moderators, etc.). We filter most of these URLs out. It is impossible to be entirely without bot accounts, but the few remaining likely do not significantly impact our data.

\begin{table}[]
\caption{Proportion of URLs that reference somewhere on Reddit and the community in which the URL is posted. The self category is percentage of URLs that link to Reddit, not total URLs.}
\label{tab:perc_reddit}
\begin{tabular}{lrr}
\toprule
category & \% reddit &  \% self \\
\midrule
makeup & 0.21 & 0.70 \\
politics & 0.36  & 0.78 \\
veganism & 0.37 & 0.51 \\
pc building & 0.38  & 0.85 \\
crafting & 0.39  & 0.33 \\
sports & 0.40 &  0.86 \\
cs & 0.41 &0.84 \\
survival & 0.44 &  0.65 \\
camping & 0.56 &  0.55 \\
%baking & 0.68 &  0.20 \\
fitness & 0.70 &  0.82\\
indoor plants & 0.76 &  0.14 \\
early retirement & 0.83 &  0.93 \\
\bottomrule
\end{tabular}
\end{table}

Unlike URLs, mentions represent talking about a community more generally than just a specific instance. In this analysis, we include Reddit-directed URLs in our mentions analysis as instances of talking about a community. It is also worth noting that mentions align closely with size, and the amount a community is mentioned is therefore generally closely aligned with the amount it mentions other communities and thus also its size.

Finally, it is worth noting that a majority of mentions are in a negative context. When we applied sentiment analysis \cite{loper2002nltk} to a sample of mention comments for every pairs of subreddits, we found mostly negative sentiment, with nearly every pair of communities having a majority negative sentiment between them. One notable exception is r/hiking as the only community where the sentiment on posts about other communities averages positive. No community averages positive sentiment on posts about them.

\subsection{Gradients of Informational and Referential Migration}
\label{sec:InformationalReferential}
We now define gradients along our informational axis (URL movement) and referential axis (mentions) in a way similar to how we define gradients for user movement. However, we do not have the same concept of entering a community that we did for users. For URLs, it is somewhat rare to be posted in the same community twice---and posting many times in a community could be considered spamming rather than genuine information diffusion. This makes our thresholding for entering a bad fit for URLs. Mentions, on the other hand, lack the concept of ``movement''. One community might be mentioned many times in another, but there is no obvious analog for the ordering of communities that a user movement has. 

\begin{figure}
    \centering
    \includegraphics[width=0.45\textwidth]{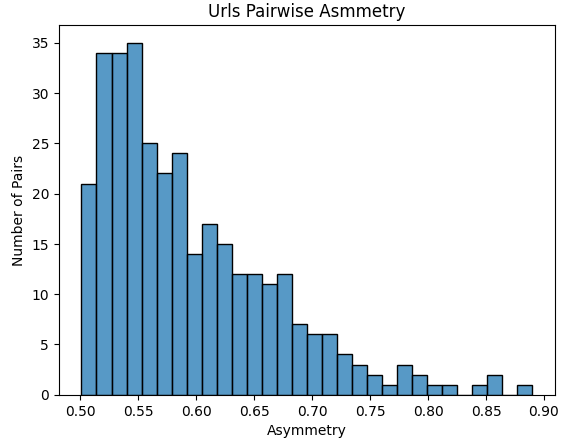}
    \caption{Histograms of the asymmetry scores for all pairs of communities by URLs.}
    \label{fig:asym_hist}
\end{figure}
\begin{figure}
    \centering
    \includegraphics[width=0.45\textwidth]{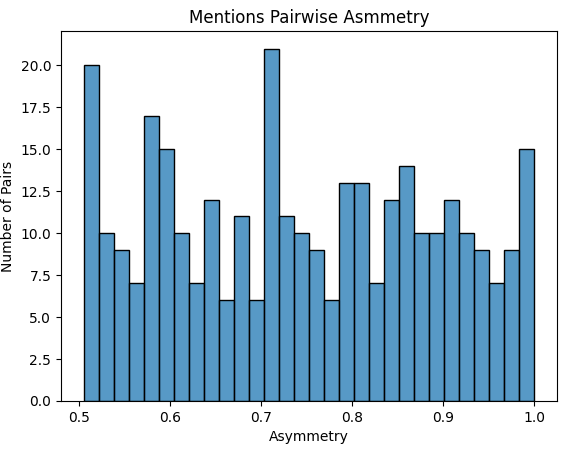}
    \caption{Histograms of the asymmetry scores for all pairs of communities by mentions.}
    \label{fig:asym_hist}
\end{figure}

Despite these differences, we can still define the orientation between a pair of communities $(A,B)$ for each of these modalities in a natural way: for information, whether a majority of URLs common to $A$ and $B$ are first posted to $A$ before $B$; and for mentions, whether there are more mentions from $A$ to $B$ than from $B$ to $A$. 
With these definitions in place, we can again define directed graphs on the communities within a category --- one for orientations based on URLs (informational), and one for orientations based on mentions (referential).
Interestingly, we again find that these graphs are acyclic, establishing the existence of informational and referential gradients in analog to our user gradients.

We can immediately see that the informational and referential measures are quite different by examining their asymmetry scores (Fig~\ref{fig:asym_hist}). URLs exhibit a bell curve shape. Mentions, on the other hand have a more uniform distribution. As with our other histograms, both measures have significantly larger asymmetry scores than expected at random. For both mentions and URLs, we restrict pair-wise analysis to pairs with at least 20 URLs/mentions between them, similar to users.

\begin{table}[]
    \centering
\resizebox{\linewidth}{!}{\begin{tabular}{llll}
\toprule
   measure & URL& Mention &Expanding\\
\midrule
  \# users& Decreases*&  Decreases&Decreases**\\
  total activity& Decreases*&  Decreases&Decreases*\\
  activity per user& Increases&  Increases&Increases\\
  toxicity & Decreases* & Decreases & Decreases** \\
  distinctiveness & Decreases* & Increases & Increases* \\
 \# URLs & Decreases*& Decreases*&Decreases*\\
 URLs per user & Increases& Decreases&Increases\\
 URL per post & Increases& Decreases*&Decreases\\
 Self-mentions& Decreases**& Increases*&Decreases**\\
 \# times mentioned& Decreases*& Increases**&Decreases**\\
 \# times mentions others& Decreases**& Decreases*&Decreases**\\
 self-mentions per post& Decreases& Decreases&Increases\\
 mentions others per post& Increases& Increases**&Increases*\\
\end{tabular}}
\caption{Binomial tests comparing source and target communities for all edges * is p-value $\leq .05$, ** is p-value $\leq .01$}
\label{tab:pairwise_binomials}
\end{table}

Given that these new axes display a similar underlying directionality to our user gradient, we examine the alignment of these new gradients. Generally, user, URL, and mention gradients do not strongly agree on directionality between pairs of communities - or even which pairs of communities have edges. Examining possible influences, however, reveals similar influences for users and URLs, but quite different influences for mentions (Tab~\ref{tab:pairwise_binomials}). 

Just as for users, URL orders seem to be strongly influenced by community size---URLs move from large subreddits to smaller ones. Given our measures of size are closely linked, users also move to communities with fewer URLs. However, this does not mean that those subreddits are more active per user, as we see no significant change in activity per user. 

Additionally, mentions are significantly influential for all of our axes. While users and URLs move to communities with fewer mentions (reflecting the size tendency), mentions direct to communities that are more self-referential. If we look at mentions relative to community size, mentions and users tend to direct to communities than mention others more, although this is not true for URLs. Users and URLs, additionally, move to communities than mention themselves less. However, mentions also seem to be influenced by URLS---mentions direct more strongly towards communities that have lower numbers of URLs per post.

Overall, this suggests that users move from larger communities to smaller ones, but also to communities that talk about others more. URLs similarly direct to smaller communities, but is not as influenced by mentions. Mentions, on the other hand, are not influenced by community size, but do direct to communities with fewer URLs that reference themselves and others more. However, none of the simple variables we check fully explain the gradient, suggesting a more complex model is necessary to explain the strong directionality.

\begin{figure}\label{fig:mentions-mags}

    \includegraphics[width=0.45\textwidth]{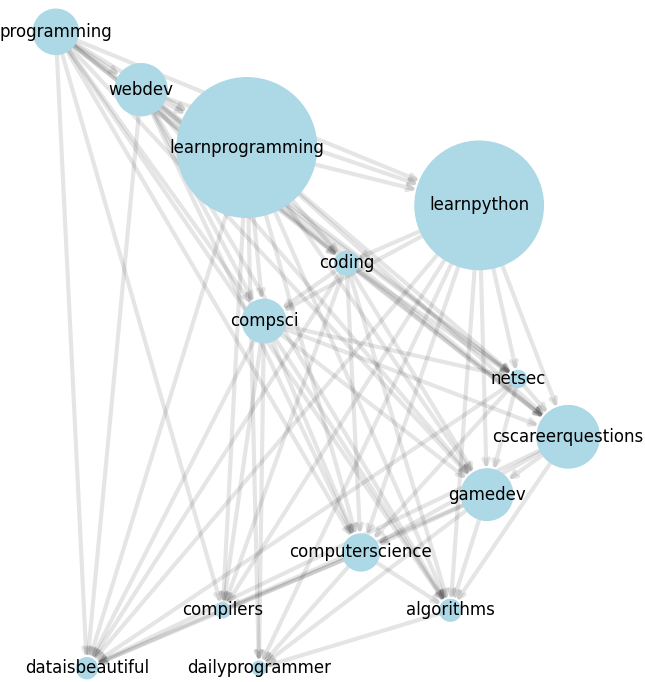}
    \includegraphics[width=0.45\textwidth]{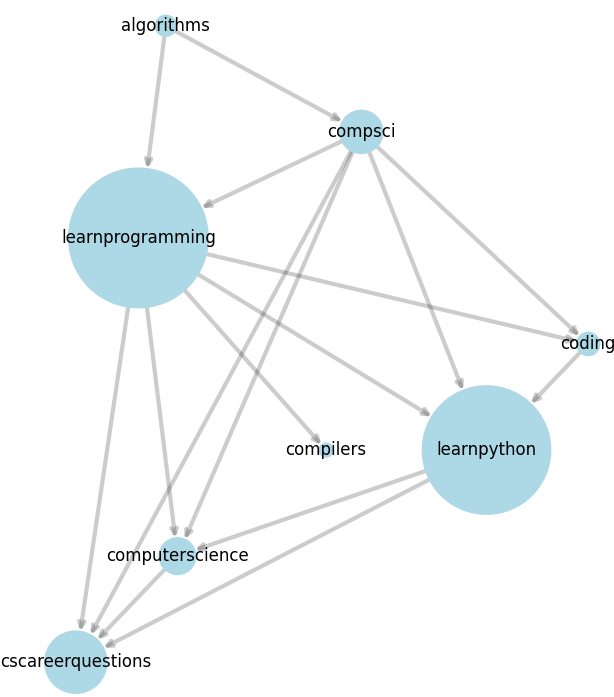}

    \caption{URL gradient graph for CS (threshold at 0.55) and mention gradient graph for CS. The thresholding for mentions to be acyclic is higher -  0.6 for CS. For all categories, thresholding is at or below 0.6 except for fitness at 0.8 and politics and makeup at 0.7. However, asymmetry of mentions is generally much higher than users or URLs, meaning higher thresholds does not strongly influence the gradient.}
    \label{fig:mentions-mags}
\end{figure}

For a visual demonstration of these gradients, we again create our small directed graphs (Fig~\ref{fig:mentions-mags}). Again, we note that these graphs are remarkably linear, although perhaps not quite as strongly so as for the user gradient. They also include different communities. While URL and user graphs seem to have similar communities across all of our categories, mentions can be very different. Additionally, URL and user gradient seem to have similar influences, while mentions has different influences. One possible nuance to this, is that sharing users or information may be seen as a measure of connectedness of communities. That is, if two communities share users or information, this may imply they are more highly connected. Mentions, however, do not necessarily fit in this framework. Mentioning another community does not imply connectedness in the same way, particularly given the generally negative sentiment.

Most of our analysis thus far has focused on the pairs of communities $(A,B)$ where the orientation is directed either from $A$ to $B$ or from $B$ to $A$.  However, as noted above, there is a third option for the orientation: 
it could be \textbf{void}. That is, there is not enough data to make an informed decision about directionality. This is particularly interesting when not all orientations are void for a particular pair of communities. For instance, Python is mentioned by many communities in the CS category, but very few users in Python post frequently enough to meet our criteria for entering, thus it does not appear in our user gradients. In fact, there is not much correlation between how many instances of our  measures exist for each gradient, with correlation less than 0.15 across the board and with mentions and URLs particularly uncorrelated at merely 0.01. 

However if we look at only zero vs non-zero indicators, then the measures seem to be slightly more aligned, with all three gradients generally agreeing on whether an edge is void. In particular, Users and URLs agree on 64\% of edges, users and mentions agree on 60\%, and URLs and mentions agree on 80\%. This is especially surprising, since if we look at only the size measures for each gradient (for a community, number of users, number of URLs, number of mentions) the sizes are much more highly correlated, since they are all correlated with the number of posts and comments for a community. However, if we compare our total sizes of each community with the gradient instances, we see a strong correlation with our user gradient ($\sim0.5$), a moderate correlation with our mention gradient ($\sim0.3$), but almost no correlation with URL gradient ($\sim0.04$). Overall, this suggests that even if the direction of an edge has limited correlation for our three measures, the existence of an edge is highly correlated.

\section{Experiments and External Data}

\subsection{Wikipedia}
\begin{figure}
    \centering
    \includegraphics[width=0.45\textwidth]{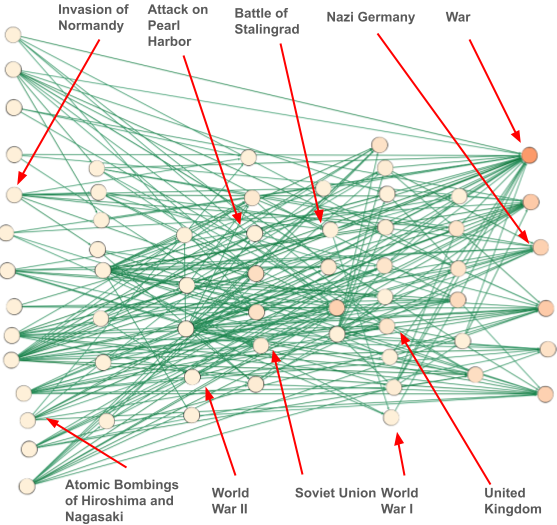}
    \caption{Graph of Wikipedia data seeded from World War II. Edges flow from left to right.}
    \label{fig:WWII}
\end{figure}

Our framework has been structured to be general enough to apply to many social media platforms. However, in order to perform analysis on Reddit, we have had to make some Reddit-specific decisions in applying the framework there -- where communities are well-defined and users typically have many interactions with a community. As an experiment, we create a similar user graph from Wikipedia (Fig~\ref{fig:WWII}). To do this, we adapt the low-level definitions to the Wikipedia context. If we consider a page on Wikipedia as a community, and any edit to the page equivalent to a comment in a community \cite{geiger2013using}, then users on Wikipedia interact far less than Reddit users and there are far fewer of them.

We first start with a seed page - say World War II. We then find all users who have edited this page (or the associated talk page), and then find all pages those editors have touched. From there, we only include pages with at least 100 editors - enough to reasonably expect overlap. Then we restrict to editors that have touched a page at least three times since 2010, and pages have connections if they have at least 30 editors in common (where at least 60\% of edits overlap in time frame). From here, we can pull a directed graph of editor movement of 62 nodes. If we threshold those edges to 0.55 we see only a few cycles and at 0.60 we have an acyclic graph of several hundred edges. This example suggests that our definitions are applicable across other social media, and the acyclicity suggests some of the findings might be similarly applicable. The same definitions applied to the seed page 'American Civil War' gave similar acyclicity results, with 42 nodes producing an acyclic graph thresholded also at 0.6.

\subsection{Simulations}
We also attempt two simulations to compare our results. We create a graph of $n$ nodes, with $m$ users. These users visit $k$ random nodes in some order, where $k$ is determined by an exponential distribution to, $f(x, \lambda) = \lambda exp(-\lambda x)$, to simulate the exponentially distributed number of communities entered by the users in our data. First, we assume there is some canonical ordering of nodes in a graph. We then assume some $p$ percentage of users follow this ordering exactly. Given this, we find that, for larger graphs, 25\% or more of users must follow this ordering in order to have reliably acyclic graphs. This could suggest a strong tendency towards ordering is required for repeated acyclicity. Note, however, that this is not a perfect representation of our data. Importantly, the histogram of the asymmetry of edges given this model is a bell curve, but it has significantly lower mean and standard deviation than our data.

We attempt a second simulation with a similar set-up, but instead using the Mallows model to give a partial ordering of nodes \cite{collas2021concentric}. Although acyclicity is reached with relatively low $\phi$, we again see a narrow, low bell curve for asymmetry values. These two results suggest that, while total ordering seems to be present to account for the acyclicity, the high levels of asymmetry suggest even higher agreement with this ordering than mere acyclicity.

\begin{figure}
    \centering
    \includegraphics[width=0.45\textwidth]{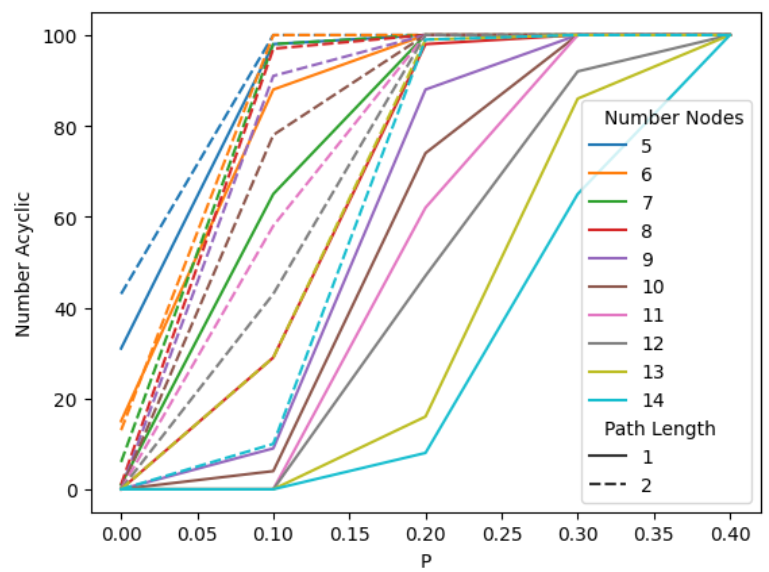}
    \caption{Number of acylic graphs from 100 simulations with some $P$ percentage of users following a prescribed ordering. Solid lines depict a path length with $\lambda = 1$, dotted is $\lambda = 2$.}
    \label{fig:user_sim}
\end{figure}

\section{Conclusion}

In this work, we have defined three types of gradients of movement within online communities---at the individual (user), informational, and referential levels.  
We developed methods to identify and analyze these gradients, and studied their properties in a broad set of topical categories on Reddit.
We find that each of these three gradients  defines an acyclic ordering to the communities, and interestingly, these orderings are not consistent with each other, with referential gradients tending to be particularly different. 

User migration is influenced by four metrics: decreasing size of the community, decreasing toxicity, increasing linguistic distinctiveness, and decreasing self-mentions/increasing other-mentions. Combined, these measures paint a picture of users moving to smaller, friendlier, and more niche communities, echoing a narrative of online specialization while seeming to dispute a narrative of radicalization. This type of progression is compatible with a number of psychological mechanisms, particularly the so-called {\em hedonic treadmill} of Brickman and Campbell, where people become satiated on their current form of an activity and seek out increasingly complex versions of it \cite{brickman1971hedonic, brickman1978lottery,diener2006beyond}. However, this raises new questions about how users find and choose communities, the interplay of all of these measures, and the dynamics of particular categories of Reddit.

Beyond these questions of mechanism, there are a number of potential further applications that could be explored using gradients.
Knowledge that gradients exist could help inform recommendations of new communities for users to explore on large platforms, and in cases where gradients potentially represent an intensification of behavior the platform views as undesirable, they could provide guidance on how to steer users away from these directions.
Given that gradients represent likely patterns of behavior in the user community, significant deviations from a gradient could also be a potential signal in the detection of bots or spam accounts.

Finally, given that we found these gradients in every location we looked - across a range of instances in both Reddit and Wikipedia - we note that they seem to be relatively pervasive across social media platforms. As such, future work focusing on meta-communities should take into account that these gradients are likely to be a baseline for any further analysis.

\bibliography{works_cited}

\newpage 

\appendix

\section{Additional Data Description}
Given our seed topics, we used Reddit's search function to identify a set of starter communities for each topic. This gave us the largest communities for any given topic. Following this, we expanded our set by utilizing the "Related Communities" sidebar feature containing other communities identified by the community moderator as having similar topics. Not all related communities were large enough for analysis and some did not align with the topic closely. We only included communities of sufficient size and specificity. Finally, we completed our community selection by identifying \textit{mentions} as described in Section \ref{sec:InformationalReferential}. For all communities that we identified for each category, we retrieved all communities mentioned by posts or comments in that community. We then included new communities that were mentioned frequently and had significant user overlap with at least one other community in the category. As before, we excluded communities that were not topical, such as large generic subreddits that have high user overlap with many communities. While there was no explicit cut-off, we aimed for about 15 to 20 communities per category. Categories with fewer communities than this did not have enough links to large, topical communities for a larger selection. In practice, this size for each category allows enough data for analysis but also ensures that all links between communities are of sufficient strength to be worth analyzing. Our final selection presents a wide array of highly related communities across the thirteen categories, which gives us a broad view of community interaction ideal for this analysis \cite{laumann1989boundary}.

While there are many other categories we could have analyzed and many other choices we could have made for communities to include, these selections represent a diverse selection of topics and demographics. This enhances the applicability of our results, demonstrating that our results are applicable to many different topics and communities (Tab~\ref{tab:cats}). It also suggests that our methods and results may be potentially applicable outside of Reddit.

Note also that some communities are older than others. For most communities, we collected data from the time of creation until Fall 2022. Two communities were banned by Reddit before then (chapotrapouse and The\_Donald, both political subreddits) and r\/nba was too large for this so we used a partial dataset containing a large fraction of years but excluding early data. For any comparative analysis, we restricted our data to only times when all communities in a category were active. For any additional analysis, the entire time frame was used.

\end{document}